\documentclass[12pt]{iopart}
\usepackage{graphicx,subcaption,dcolumn,bm,amssymb}
\usepackage{hyperref}
\usepackage[dvipsnames]{xcolor}
\usepackage{soul}
\definecolor{purp}{HTML}{832591}
\DeclareCaptionJustification{justified}{\leftskip=0pt \rightskip=0pt \parfillskip=0pt plus 1fil}
\hypersetup{
  colorlinks,
  allcolors = {purp},
  citecolor = {purp},
  allbordercolors = {white},
}
\newcommand{\sabq}{\ensuremath{S^{\alpha \beta}(\mathbf{q})}}
\newcommand{\slq}{\ensuremath{S^{L}(\mathbf{q})}}
\newcommand{\stq}{\ensuremath{S^{T}(\mathbf{q})}}

\begin{document}

\title{Electric field fluctuations in the two-dimensional Coulomb fluid}

\author{Callum Gray$^{1,2}$, Steven T. Bramwell$^2$ and {Peter C. W. Holdsworth}$^3$}
\address{$^1$ School of Biological and Chemical Sciences, Queen Mary University of London, Mile End Road, London E1 4NS}
\address{$^2$ London Centre for Nanotechnology and Department of Physics and Astronomy, University College London, 17\textendash{}19 Gordon Street, London WC1H 0AH, United Kingdom}
\address{$^3$ Universit\'e de Lyon, ENS de Lyon, Universit\'e Claude Bernard, CNRS, Laboratoire de Physique, F-69342 Lyon, France}
\ead{callum.gray@qmul.ac.uk}

\begin{abstract}
The structure factor for electric field correlations in the two dimensional Coulomb fluid is simulated and compared to theories of the dielectric function. Singular changes in the structure factor occur at the BKT insulator to conductor transition, as well as at a higher temperature correlation transition between a poor electrolyte and perturbed Debye-H\"uckel fluid. Structure factors are found to differ in the canonical and grand canonical ensembles, with the poor electrolyte showing full ensemble inequivalence. We identify mechanisms of `underscreening' and  `pinch point' scattering that are relevant to experiments on ionic liquids and artificial spin ice respectively. 
\end{abstract}

\maketitle

\section{Introduction}

Field correlations in two- and three-dimensional Coulomb fluids are relevant to many processes in physics, chemistry and biology~\cite{Oosawa}. The theory of them goes back many years~\cite{Oosawa, ZH, EK}, but direct numerical and experimental tests were lacking. More recently, the development of local algorithms for electrolytes~\cite{MR, Faulkner} and experiments that image field correlations in emergent Coulomb fluids~\cite{Fennell, Chang, Perrin, Ostman} have provided opportunities for progress, while studies of anomalous response and screening in ionic liquids~\cite{Weingartner, Perkin} motivate an urgent reappraisal of the many-faceted theory. In this context, a numerical, field-theoretic attack on the problem is timely and the two dimensional lattice Coulomb fluid -- rich, accessible and with a Berezinskii-Kosterlitz-Thouless (BKT) confinement-deconfinement transition~\cite{SP,Berezinskii,KT,Kosterlitz} --  is the logical place to start~\cite{LevinCamp}.




In a primitive model of charges $\pm Q$, the evolution of the dielectric function and implicitly, the structure factor for field correlations, through the BKT transition~\cite{Berezinskii,KT,Kosterlitz} temperature, $T_{\rm KT}$, was discussed in early works by Zittartz and Huberman~\cite{ZH} (ZH) and Everts and Koch~\cite{EK} (EK). ZH treated the low density limit
where the BKT transition manifests as a singularity in the pressure function at $\frac{Q^2}{8\pi \epsilon_0 k_BT}=1$~\cite{SP,ZH} and 
predicted that the conducting phase is divided into two regimes by a second temperature, $T_2 =2T_{\rm KT}$. In the temperature interval $T_{\rm KT} <T< T_2$, ZH's `poor electrolyte' regime, the logarithmic Coulomb interaction between charges leads to singularities in the partition function, ensuring scale free behaviour over a divergent ‘inertial range’ of length scales,  including the ultraviolet cut off, $a$. The classical response of a standard electrolyte, which can be described by Debye-H\"uckel theory and its corrections, only appears above $T_2$, marking a `correlation transition'~\cite{Daruka}.

EK generalised this work to finite density, showing that the inertial range is eventually cut off at large scale by a screening length which itself is a non-analytic function of density. These simple arguments were confirmed by mapping to the Sine-Gordon equation~\cite{Minnhagen} and using renormalisation techniques~\cite{Young, foot}. All calculations suggest that, while $T_{\rm KT}$ is shifted at finite density, $T_2$ is density independent but with reduced inertial range as density increases. Using the units of Ref. \cite{LeeTeitel}, with Boltzmann's constant $k_{\rm B}=1$, free space permittivity $\epsilon_0 = 1/2\pi$ and charge $Q = \pm 1$,  gives the upper limit for $T_{\rm KT}=\frac{1}{4}$ and that for $T_2=\frac{1}{2}$.

We have simulated the static structure factor for electric field correlations of a two dimensional lattice Coulomb fluid across its rich phase diagram~\cite{LeeTeitel,GuptaTeitel}. 
Our model is described below, but in brief, 
we apply the algorithm described in detail in Refs. \cite{MR,Faulkner} and summarised in Appendix A. In this paper we present results with the single particle core energy set to zero, a situation compatible with magnetic systems and superfluids~\cite{KT}. Two dimensional electrostatics, with definitions of the chemical potential and core energy, are summarised in Appendix B. The system sizes for the present study are $L^2$ on a square lattice, with $L \le 256$ and the lattice constant $a$ taken to be unity. Zero core energy corresponds to fugacity, $z=\exp(\beta\mu^{2D})$, with $-2\mu^{2D}$ the purely electrostatic energy cost of introducing an isolated neutral pair of charges, separated by the lattice parameter \cite{KT}.  This gives a small but non-zero value of $z$ which reduces the BKT transition to $T_{\rm KT}=0.215$~\cite{Faulkner}, but for which the unbinding picture remains valid.

In the rest of the paper we introduce the relevant theory (section 2) before going on to test our simulated structure factor against those predicted by ZH and EK (section 3). We demonstrate full consistency with the EK theory, including the existence of a well-defined correlation singularity at $T_2$. We do not address the thermodynamic consequences of this transition but we note that it has been discussed in the literature~\cite{Young, Gallavotti, Fisher-Levin}. At the level of the structure factor, we find that the poor electrolyte is further characterised by a breakdown of ensemble equivalence between the canonical and grand canonical ensembles, a consequence of the divergent inertial range of the contributing length scales~\cite{Campa}. Such effects are striking signatures of the approach to topological order~\cite{KT}, but they would be challenging to study experimentally in BKT systems such as magnets and superfluid films~\cite{KT}. Therefore, in our discussion (Section 4), we briefly considering the relevance of our results to more accessible systems such as ionic liquids~\cite{Weingartner, Perkin} and artificial spin ice~\cite{Perrin, Ostman}.


\section{Theory}

The classical Coulomb fluid is an system of interacting electric charges which is overall charge neutral. The Coulomb energy of the system may be considered to be stored in the local electric field ${\bf E}({\bf r})$, the charges being topolological defects in that field (that is, defects that cannot be removed by stretching or bending the field lines).  The energy may be elegantly expressed in terms of an integral over the field squared, constrained only by Gauss' law. 
\begin{equation}\label{fluid}
U = \frac{\epsilon_0}{2} \int |{\bf E}|^2\, d^3 r~~~~~~~\nabla \cdot {\bf E} = \frac{\rho}{\epsilon_0}
\end{equation}
where $\rho({\bf r})$ is the local charge density. This returns Coulomb's law of interaction between charges, which goes as $1/r$ in three dimensions and logarithmically with $r$ in two dimensions. 
The most general solution of Gauss' law shows that ${\bf E}$ has both irrotational and solenoidal components~\cite{fh}. The irrotational component is the ordinary field of electrostatics, the negative gradient of a scalar potential.  The solenoidal component is generally discarded in electrostatic problems, but more generally can be thought of as corresponding to transverse photon-like degrees of freedom: closed field loops with no sources. We retain this component as an object of interest in its own right. To do so does not compromise the electrostatic description because the energies of the irrotational and solenoidal fields are additive (which follows from Eq. \ref{fluid}) and, in the absence of electrodynamic coupling, the partition function factorises.

The model we treat is a lattice version of this~\cite{MR,Faulkner}, where the integral becomes a sum and $\nabla$ becomes a lattice divergence operator~\cite{Faulkner}: 
\begin{equation}\label{lattice}
U = \frac{\epsilon_0 a^2}{2} \sum_i |{\bf E}_i|^2~~~~~~~\nabla \cdot {\bf E} = \frac{\rho}{\epsilon_0}.
\end{equation}
Specifically, we treat the primitive model of a symmetric two dimensional Coulomb fluid on a square lattice. The fields ${\bf E}$ are situated on the lattice bonds and the charges are introduced on the vertices of the lattice by means of a chemical potential  $\mu^{\rm 2D}$ (for details of how this is defined, see Appendix B). As we specialise to the case of single charges only, with zero core energy, the charges cost energy only insofar as they alter the local fields~\cite{Faulkner}. In this grand canonical representation, the number density of charges $n(t)$ (defined below) will of course vary with temperature, going to zero at low temperature and saturating in the high temperature limit. Details of our simulations are given in Appendix A.  

Field correlations are most clearly visualised in reciprocal space. The irrotational and  solenoidal field components Fourier transform, respectively, to longitudinal (L) and transverse (T) components, ${E}^{\rm L}$ and ${E}^{\rm T}$, which fluctuate independently in the electrostatic limit.  The corresponding structure factors, $S^{\rm L} = \langle E^{\rm L}({\bf q})E^{\rm L}(-{\bf q})\rangle $ and $S^{\rm T}= \langle E^{\rm T}({\bf q})E^{\rm T}(-{\bf q})\rangle $, are the eigenvalues of the structure factor tensor $S^{\alpha \beta}({\bf q})$. These eigenvalues are periodic with the reciprocal lattice $\{{\bf G}\}$ as shown in Fig.1a. 

Of most interest is the longitudinal structure factor $S^{\rm L}$ as this characterizes the fluctuations of the irrotational electric fields that emanate from the charges in the system. It is related to the Fourier transform of the charge-charge correlation function via Gauss' law: 
\begin{equation}
  S^{\rm L}(q) =\frac{a^2}{-\epsilon_0^2{\bigtriangleup_q}} \langle \rho({\bf q})\rho(-{\bf q})\rangle,   
\end{equation}
where $\rho({\bf q})$ is the Fourier transform of the local charge density and where 
\begin{equation}
\bigtriangleup_q = 2 -  \cos(q_x a) - \cos(q_y a)
\end{equation}
is the lattice Laplacian which reduces to the $-q^{2}a^2$ expected of continuous systems at long wavelength. 

\begin{figure}[htb]
  \centering
  \begin{subfigure}{0.5\textwidth}
    \centering
    \includegraphics[width=\textwidth]{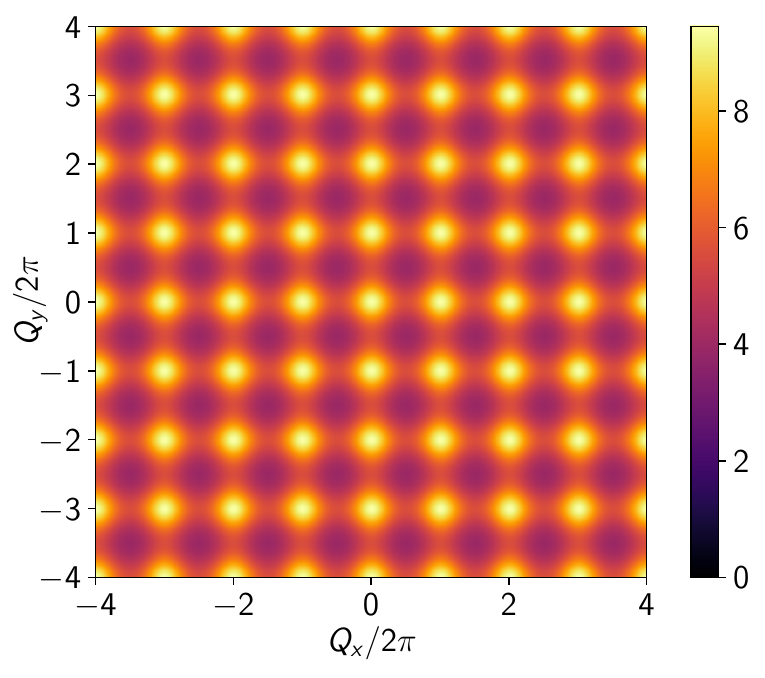}
    \caption{}
  \end{subfigure}
  
  \begin{subfigure}{0.49\textwidth}
    \centering
    \includegraphics[width=\textwidth]{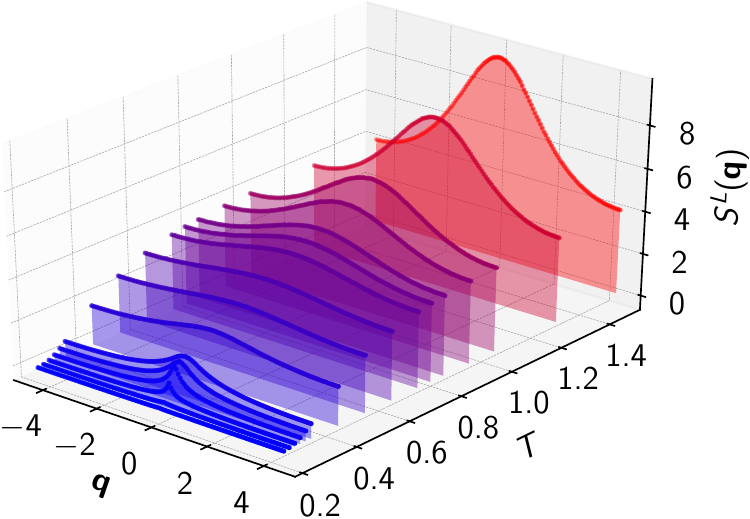}
    \caption{}
\end{subfigure}
  \begin{subfigure}{0.49\textwidth}
    \centering
    \includegraphics[width=\textwidth]{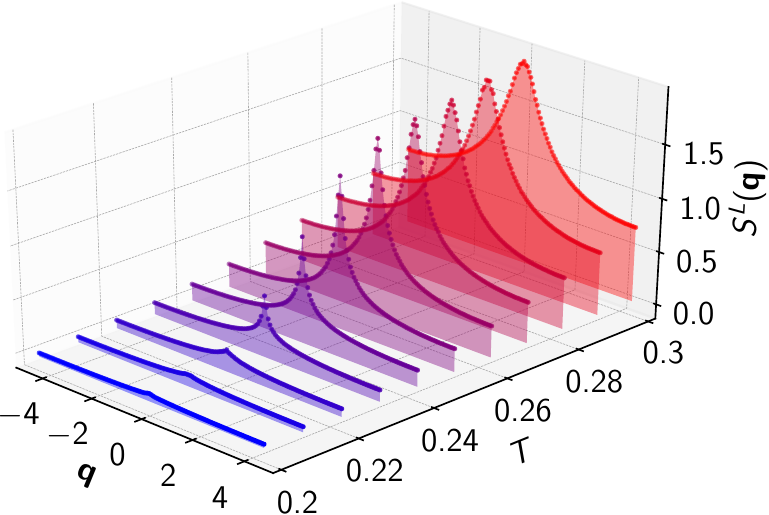}
    \caption{}
  \end{subfigure}
\caption{(a) Longitudinal structure factor $ S^{\rm L}(\mathbf{Q})~(T = 1.5, L=128)$ across several Brillouin zones. (b) Line shape of $S^{\rm L}(q)$ along the trajectory $[1,1]$ showing a non-monotonic evolution with temperature (here ${\bf q = Q-G}$ where ${\bf G}$ is a reciprocal lattice vector). (c) Zoom in to show a cusp gradually forming near to the BKT transition at $T_{\rm KT}=0.215$.}
\end{figure}

Figure 1b shows the thermal evolution of $S^{\rm L}(q)$, sweeping over a large temperature range. Empirically, the structure factor is found to rapidly narrow in $q$ and tend to a multi-Lorentzian form at $T \lesssim T_2$ ~\cite{Callum}. A finite cusp forms below $T_{\rm KT}$ that tends to diverge above the transition (Fig. 1c) and is finally rounded above $T=0.28$. Rather surprisingly, on heating well above the transition, the line shape evolves in a non-monotonic way: it first flattens and then sharpens again. 

A quantitative analysis of the lineshape may be achieved by relating the structure factor to the electrostatic susceptibility and to the static dielectric function. The susceptibility is the response of the internal field to an external field ${\bf D}$,

\begin{equation}
\chi(q)  = - \frac{\epsilon_0 {\bf E}^{\rm L} -{\bf  D}}{\bf D}, 
\end{equation}
which is related to the structure factor for field correlations: 
\begin{equation}
S^{\rm L}(q) =a^2 kT \chi(q)/\epsilon_0
\end{equation}
and to the dielectric function by 
\begin{equation}
\chi(q) = 1-\frac{1}{\epsilon_q}~~~\Rightarrow~ ~~\epsilon_q = \frac{1}{1- \chi(q)}.
\end{equation}
This can be written in Dyson form 
\begin{equation}
\epsilon_q=1+\chi(q)\epsilon_q=1+\frac{1}{-k_BT\epsilon_0{\bigtriangleup_q}} \langle \rho({\bf q})\rho(-{\bf q})\rangle\epsilon_q,
\end{equation}
and developed perturbatively in diagrammatic series.

This is the approach taken by EK \cite{EK} in the low charge density limit, where the small parameter is the fugacity $z$. For systems with short ranged charge correlations, and number density of charges $n = n_++n_-$, EK show that $\langle \rho({\bf q})\rho(-{\bf q})\rangle= (1+F) n a^2 Q^2$, with $F$ a constant of order unity, as all but short ranged off-diagonal terms in the correlation function sum to zero. 
This yields the following form for the longitudinal structure factor:
\begin{equation}\label{norm}
S^{\rm L}({\bf q}) = \frac{a^2 k_{\rm B}T}{\epsilon_0}\frac{\kappa^2 a^2 (1+F(T))}{\kappa^2 a^2 (1+F(T))+(-\Delta_q)}
\end{equation}
The limit of weak correlations, in which $F=0$, corresponds to the Debye-H\"uckel dielectric function $\epsilon_q =1-   \kappa^2 a^2/\bigtriangleup_q$, where $\kappa  = \sqrt{n Q^2/\epsilon_0 kT}$ is the reciprocal Debye length, and hence the Debye-H\"uckel structure factor
\begin{equation}\label{DH}
S^{\rm L}_{\rm DH}({\bf q}) = \frac{a^2 k_{\rm B}T}{\epsilon_0}\frac{\kappa^2 a^2 }{\kappa^2 a^2+(-\Delta_q)}.
\end{equation}
 EK find that this logic is satisfied for $T>2T_{\rm KT}$ with $F(T)$ a positive, temperature dependent function falling to zero at high temperature. 

However, in the poor electrolyte regime, $T_{\rm KT}<T<2T_{\rm KT}$, the inverse screening length in Eq. \ref{DH} must be replaced by the non-analytic function~\cite{EK}
\begin{equation}
\kappa\rightarrow \tilde{\kappa}= C(T)n^{\nu/2}, \;\; \nu=\frac{1}{2}\left(\frac{T}{T-T_{\rm KT}}\right),\label{nu}
\end{equation}
such that $\tilde{\kappa}^{-1} \gg \kappa^{-1}$ throughout. Accordingly, the structure factor is predicted by EK~\cite{EK} to follow an anomalous law at $q>\tilde{\kappa}$,  
\begin{equation}\label{anom1}
S^{\rm L}\sim\left( \frac{q}{\tilde{\kappa}}\right)^{-2/\nu}~~~~~~~~
\frac{q}{\tilde{\kappa}}>1,
\end{equation}
crossing over to classical behaviour for $q$ below this anomalously small threshold:
\begin{equation}\label{anom2}
S^{\rm L}\sim \frac{\tilde{\kappa}^2}{q^2+\tilde{\kappa}^2}~~~~~~~~~ 
\frac{q}{\tilde{\kappa}}< 1.
\end{equation}
The crossover at very small $q$ ensures that this is not a critical regime and does not have singular thermodynamic measures, except at $T_{\rm KT}$. 
 
As mentioned in the introduction, $T_{\rm KT}$ is shifted for finite charge density, from $T_{\rm KT} = 0.25$ to $T_{\rm KT} = 0.215$~\cite{Faulkner}, but Eq. (\ref{nu}), as written, applies to zero charge density, yielding $\nu = 1$ at $T_2 = 0.5$.  We conjecture that the correct form for finite charge density is Eq. (\ref{nu}) but with $T_{\rm KT}$ set to its renormalised value, 0.215.  We test the validity of this conjecture near $T_{\rm KT}$ by making quantitative analysis of our data. 
\begin{figure}[htb]
  \centering
  \captionsetup{justification=justified,singlelinecheck=off}
  \begin{subfigure}{0.45\textwidth}
    \includegraphics[width=\textwidth]{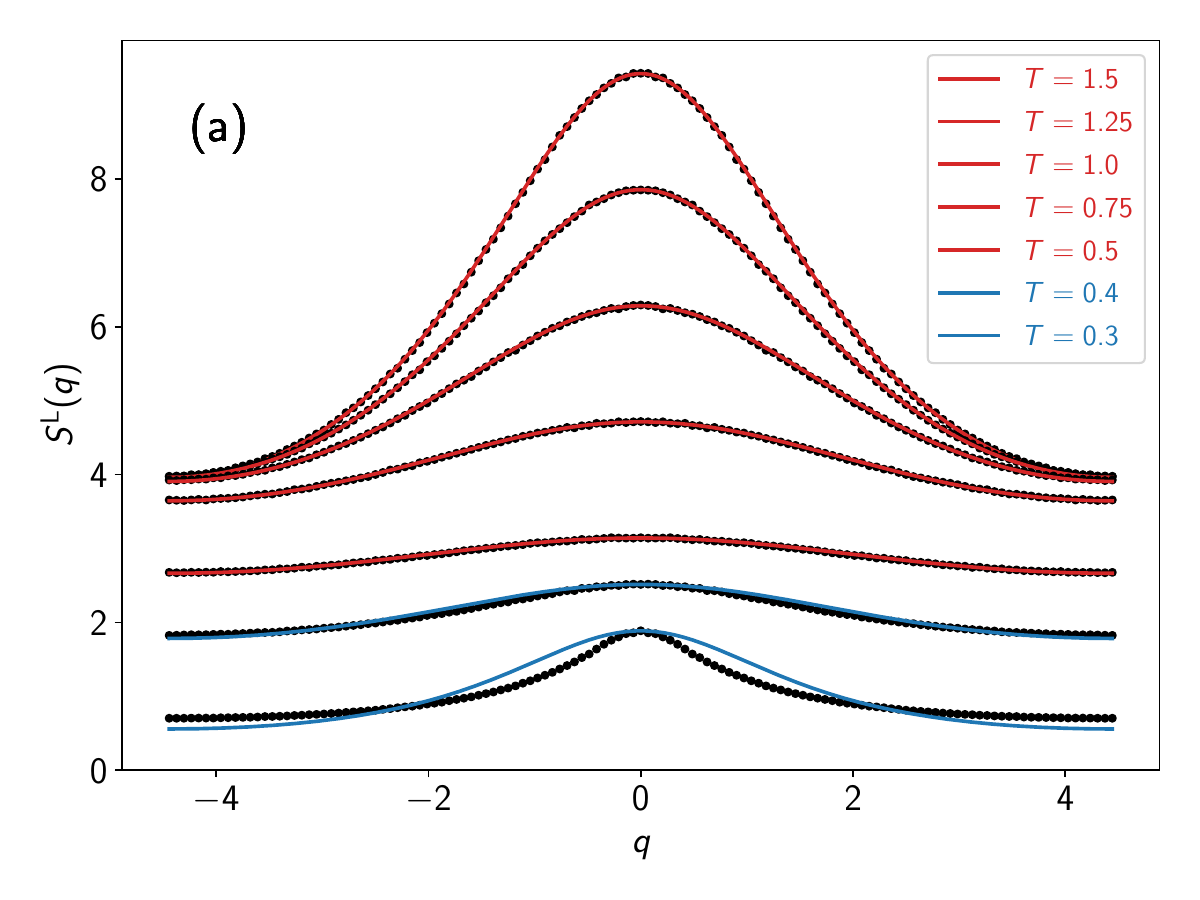}\label{}
  \end{subfigure}
  \begin{subfigure}{0.48\textwidth}
    \includegraphics[width=1.12\textwidth]{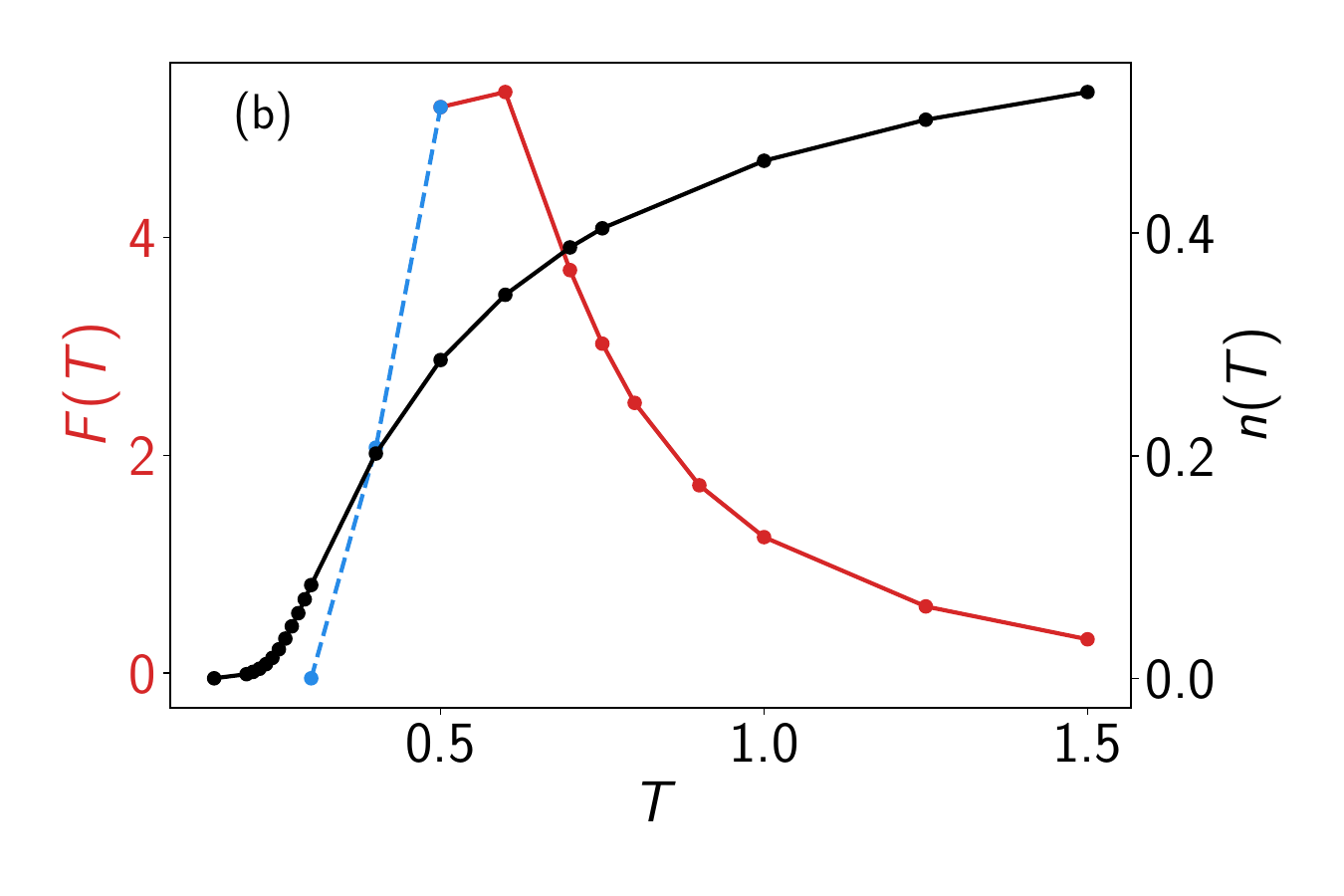}\label{}
\end{subfigure}
\caption{{How the structure factor, charge density $n(T)$ and parameter $F(T)$ that corrects the screening length vary with temperature.  (a) Simulated structure factor ($L=128$, black points) versus Eq.(\ref{norm}), i.e. 
$
S^{\rm L}({\bf q}) = (a^2 k_{\rm B}T/\epsilon_0)\eta^2/\left(\eta^2+(-\Delta_q)\right)
$ 
with $\eta^2=\kappa^2 a^2 (1+F(T))$~\cite{EK} (top-bottom order of curves matches the order of temperatures given in the inset). Red, blue lines indicate the standard and poor electrolyte regimes respectively. (b) 
Fitted $F(T)$ (same colour code) and simulated density $n(T)$ (black, lines are guides to the eye.)}}
\end{figure}

In summary our simulated data needs to be compared with Eq.(\ref{norm}) at $T> T_2$ and Eqs.(\ref{nu},\ref{anom1},\ref{anom2}) in the approach to $T_{\rm KT} = 0.215$.    

\section{Results}

Our results are summarised in Figs. 2, 3, where we show that the simulated $S^{\rm L}(q)$ can be divided into the three temperature regions. 
In {\bf regime (i)},  $T> 0.5$, 
using $F(T)$ as a fitting parameter, we find near-perfect agreement between EK theory and simulation (Fig.2a).
 The best fit value of $F(T)$, shown in Fig. 2b increases from zero at high temperature and appears to diverge as $T=0.5$ is approached from above, confirming that there is indeed a singular change in the form of the structure factor at, or near this temperature. The surprising sharpening of the line shape at high temperature arises because  $F \rightarrow 0 $ with increasing $T$ and $n(T)$ saturates (Fig. 2b), so the line shape sharpens as $1/\sqrt(T)$ and the system becomes a dense electrolyte described quantitatively by Debye-H\"uckel theory. A finite size scaling analysis (Fig. 4, upper) reveals near-perfect data collapse, showing that the screening length is well below the simulated scales.

In {\bf regime (ii)}, the poor electrolyte at $0.215 < T < 0.5$, the single function $F(T)$ does not fit the data and the analytic EK function progressively fails below $T=0.5$, with the expected crossover between anomalous response and a quadratic regime at small $q$ becoming visible (Fig. 2a). Figure 3 shows how, just above the shifted $T_{\rm KT}$,
the EK form (Eqn. \ref{nu}), which predicts small exponents in the range  $2/\nu = 0.09 - 0.56$ on the Figure, is fully consistent with our data. This also supports the proposed shift in $T_{\rm KT}$.

\begin{figure}[ht]
  \centering
    \includegraphics[width=0.6\textwidth]{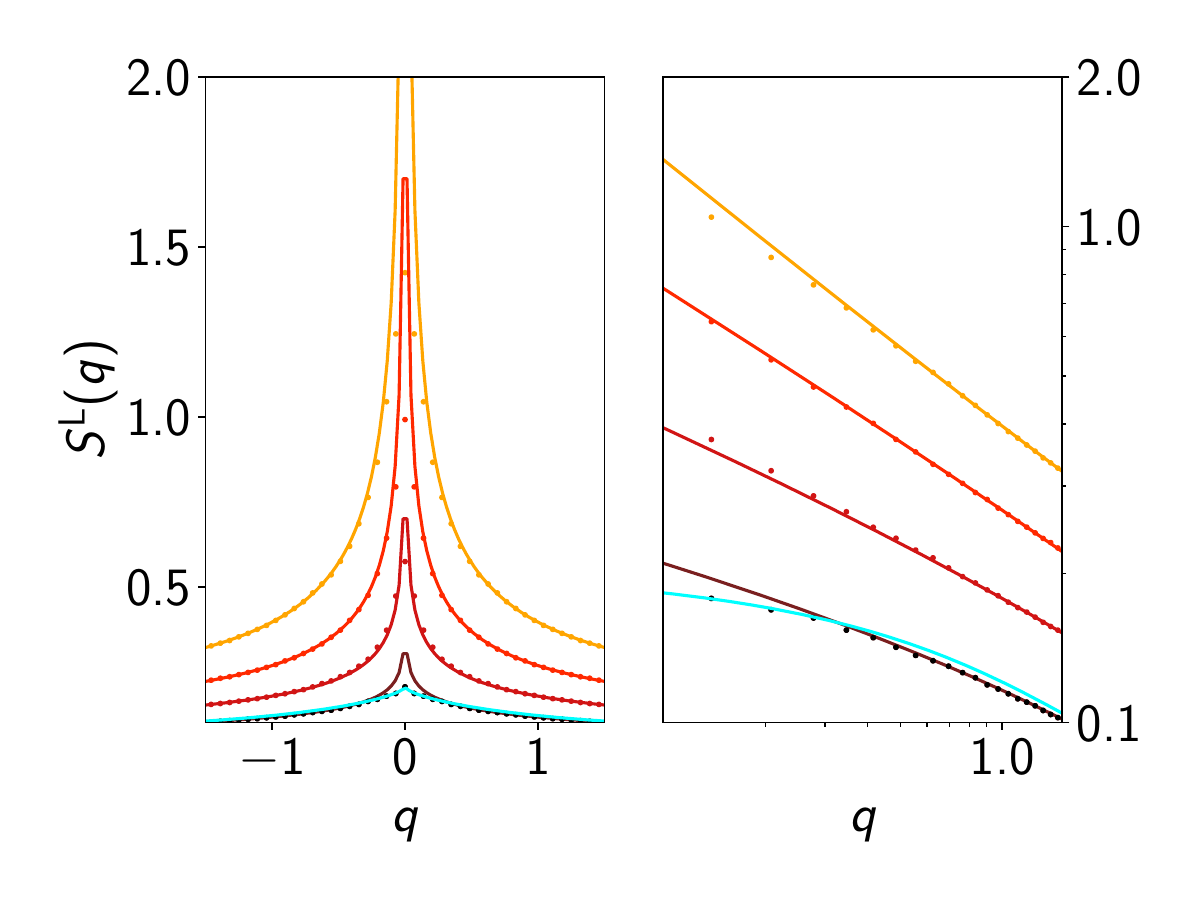}\label{}\vspace{-0.25cm}
\caption{{EK form $S^{\rm L}\sim\left( \frac{q}{\tilde{\kappa}}\right)^{-2/\nu(T)}$ (lines) versus 
simulated data ($L=128$, points) at $T = 0.22$, $0.23$, $0.24$, $0.25$ (bottom\textendash{}top). 
Lines (except cyan) are $A(T) + B(T) |q|^{-2/\nu(T)}$ where $A, B$ are determined by fitting at $q = 0.75,1.25$. Cyan line is the rescaled ZH function (see Appendix C). Scales are natural (left) and logarithmic (right).}}
\end{figure}


A finite size scaling analysis (Fig. 4 and Appendix D) shows the finite-$q$ power law regime to have small finite size corrections going as $1/L$,
but the small-$q$ quadratic regime to have much larger power law corrections with small exponents of the order $1/\nu$. This makes any approach to the thermodynamic limit impossible in our finite simulations for $q\rightarrow 0$.  Hence, while we expect the Stillinger--Lovett sum rule~\cite{SL} for deconfined charge, equivalent here to $\lim_{q \to 0} S^{\rm L}(q) = 2\pi T$, to apply in the thermodynamic limit for all $T > T_{\rm KT}$, the strong finite size effect precludes its observation in regime (ii).

In {\bf Regime (iii)}, $T<T_{\rm KT}$, we find a finite cusp singularity at $q=0$. ZH provide a closed form for the structure factor in this regime (reproduced in Appendix C) which fits the data with a single fitting parameter (see Appendix C). A scaled ZH form, appropriate to regime (iii), also describes the observed cusp at $T = 0.22$, just above $T_{\rm KT}$ (see Fig. 3), consistent with the expected shift in the BKT transition in a finite system, which varies logarithmically with system size \cite{BH}.
  
\begin{figure}[htb]
  \centering
    \includegraphics[width=0.6\textwidth]{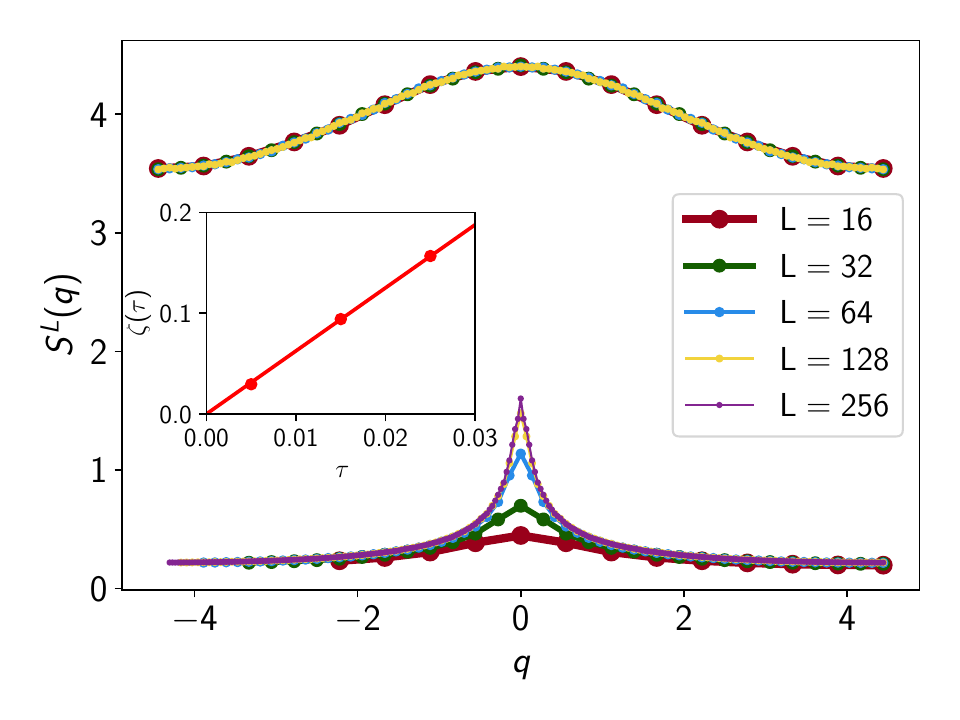}\vspace{-0.25cm}
\caption{{Size dependence of data at (upper)  $T=0.7$ (line is EK form) and (lower) $T = 0.23$ (lines are guides to the eye). Inset: fitted exponent $\zeta(\tau)$ in $S(0) = 2 \pi T - \alpha(T) (1/L)^{\zeta(\tau)}$}, where $\tau = T-T_{\rm KT}$ and $T_{\rm KT} = 0.215$.}
\end{figure}

One of the consequences of long range interactions is the possibility of ensemble inequivalence~\cite{Campa}. In the case of Coulomb interactions, screening typically regularises the interactions ensuring ensemble equivalence for thermodynamic variables.  However, even in this case, structure factors could show differences at finite wavevector. The results of our preliminary investigations of this question are shown in Fig. 5. where we compare simulated data in the two ensembles with the canonical density tuned to the grand canonical average at fixed $z$. In the classical electrolyte regime, for $T=1$ where we previously fitted data with $F\approx 1.5$, we find a considerable difference for the canonical structure factor. It can be fitted over a large range of $q$ with the Debye-H\"uckel function, $F=0$, coinciding at $q=0$ and appearing to cross over back to the grand-canonical function for large $q$. 

At $T=0.3$, in the poor electrolyte regime, the canonical structure factor is much narrower and of smaller amplitude over the entire Brillouin zone, including $q=0$. This result gives a hint of complete thermodynamic ensemble inequivalence in this intermediate regime. It suggests that, although the result at  zero density, $T_{\rm KT}=0.25$ is surely ensemble independent, the renormalisation of $T_{\rm KT}$ at finite charge density may not be. A detailed analysis of this question is beyond the scope of the present work but could be the subject of future studies. 
\begin{figure}[htb]
\centering
    \includegraphics[width=0.6\textwidth]{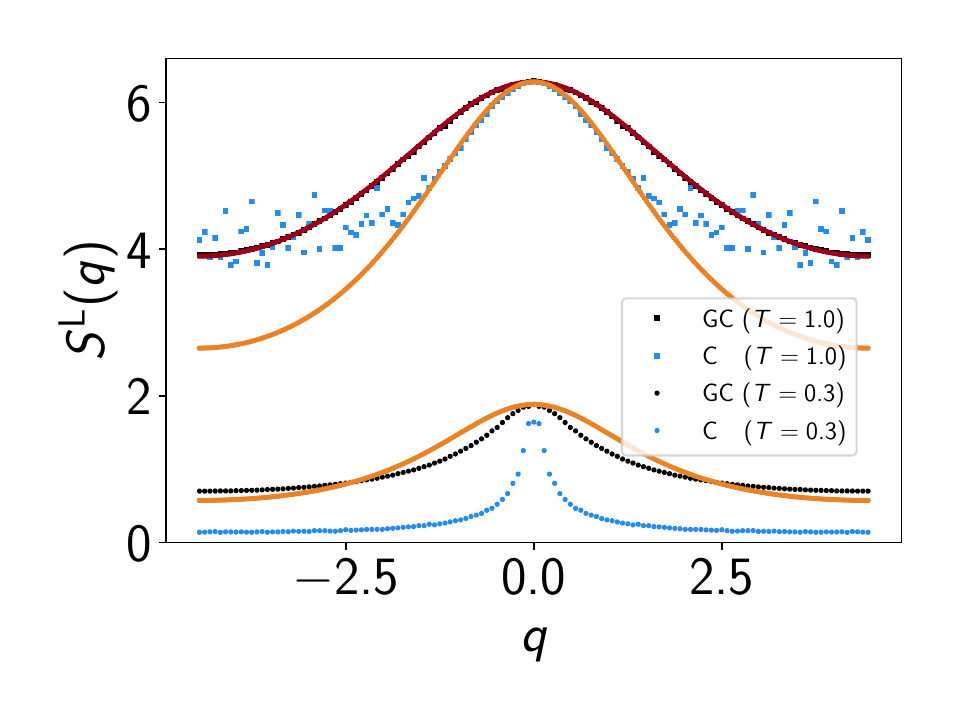}\label{}\vspace{-0.25cm}
    \caption{{Grand canonical ensemble (GC) and canonical ensemble (C) results at two $\{ T,n \}$ combinations ($L=128$, red line $=$ fit. of Fig.2a), compared with the Debye-H\"uckel prediction (orange line).}}
\end{figure}

\section{Discussion}

Having described our main results, we conclude the paper by commenting on their relevance to two particular
experiments: on `underscreening' in ionic liquids~\cite{Perkin} and on superspin correlations in artificial spin ice~\cite{Perrin,Ostman}. 

The term `underscreening' implies a screening length that is greater than the Debye length. High density ionic fluids in confined geometry appear to be strongly underscreened~\cite{Perkin}, as does a model two-dimensional Coulomb fluid of point particles~\cite{Samaj2002}, but the origin of this remains an open question. In any real ionic fluid, the dielectric function, and hence field correlations at large $q$, will depend on local chemical details, or the precise short-ranged form of the potential~\cite{Xiao}, but long ranged underscreening is more likely a generic property that can be captured by our model \footnote{Here, the different form of the Coulomb interaction in two and three dimensions (logarithmic versus $1/r$ respectively) will lead to quantitative differences, but, we argue, not qualitative ones: the field theoretic hamiltonian and reciprocal space interaction ($1/q^2$) are the same in both cases and will mediate qualitatively similar effects.}.
Referring to our results, we first note that the normal electrolyte is `overscreened' (Fig. 2): that is, the inverse screening length is  $\kappa \sqrt{1+ F(T)}$, where $F(T)$ is positive, so the screening length is generally shorter than the Debye length, $1/\kappa$. Despite this, our analysis of EK theory does reveal two mechanisms for underscreening.  
First, the poor electrolyte becomes massively underscreened (see Eq.\ref{nu}) as $\tilde{\kappa}$ diverges in the approach to $T_{\rm KT}$ from above.  Here, the increasing formation of multi-scale dipoles reduces the effective free charge concentration and frustrates the screening to expose the long ranged interaction (Fig. 3). Second, a restriction on channels for particle exchange tends to enhance the screening length, as evidenced by the ensemble inequivalence we have found (Fig. 5, where the canonical ensemble is underscreened).  Both mechanisms may be broadly relevant to ionic liquids: the first, because dipolar correlations can be important in three dimensional Coulomb fluids despite there being no BKT transition, and the second because any loss of ergodicity with respect to particle exchange could mimic ensemble inequivalence in this regard. 


Turning now to comparison with spin ice, the magnetic correlations in these materials show striking `pinch point' patterns in the neutron scattering cross section~\cite{Fennell}. These patterns are a diagnostic of the field correlations of the emergent electromagnetism specific to systems with discrete translational symmetry. In general a scattering pattern is a projection of the structure factor tensor; pinch points can result when the tensor is anisotropic. The structure factor tensor $S^{\alpha \beta}({\bf q})$ is periodic in the reciprocal lattice vector ${\bf G}$, while the differential cross section of magnetic neutron scattering, for example,  measures its projection, transverse to the scattering vector ${\bf Q}= {\bf G} + {\bf q}$.  The result is that any  anisotropy in $S^{\alpha \beta}({\bf q})$ is repeated in an aperiodic pattern as ${\bf Q}$ crosses Brillouin zone boundaries. In the Coulomb phase of spin ice, in which monopoles are absent, the emergent fields are solonoidal so that the only finite components of $S^{\alpha \beta}({\bf q})$ are perpendicular to ${\bf q}$, resulting in a highly anisotropic tensor and sharp pinch point singularities in the scattering pattern. The square lattice symmetry studied here can be found in artificial spin ice metamaterial arrays~\cite{ASIreview}. Although the length scales of the micromagnetic elements (superspins) are beyond those accessible for neutron scattering, effective neutron scattering cross sections can be constructed from the Fourier transform of the superspin correlation function, measured in direct space with the appearance of analogous square pinch point patterns~\cite{Perrin, Ostman}.

\begin{figure}[htb]
  \begin{minipage}[h]{.39\linewidth}
  \centering
  \begin{subfigure}{\linewidth}
  \vspace{-2cm}
    \includegraphics[width=\textwidth]{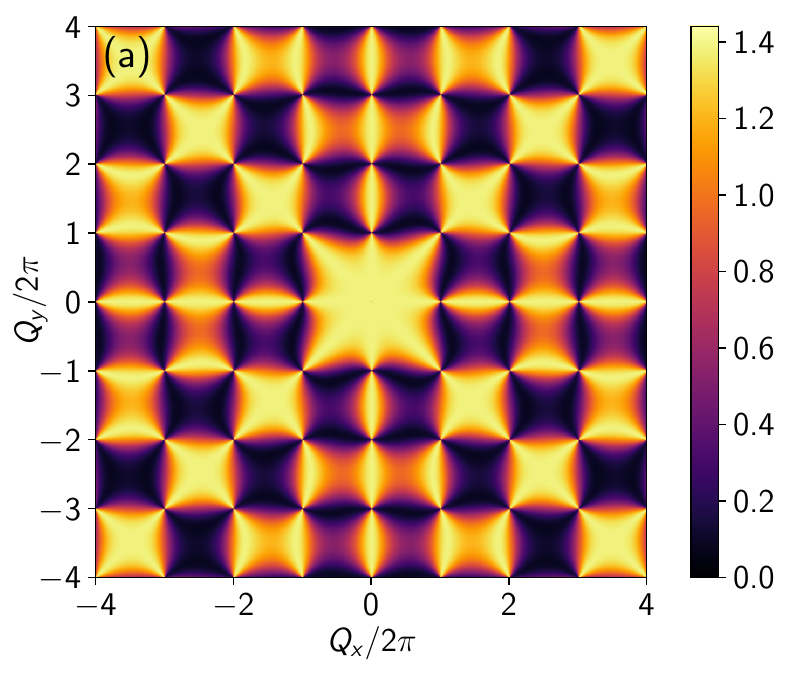}\label{sfig:sperp_t020}
\end{subfigure}
\end{minipage}
\begin{minipage}[b]{.2\linewidth}
  \centering
  \vspace{0.25cm}
  \begin{subfigure}{0.8\linewidth}
    \includegraphics[width=\textwidth]{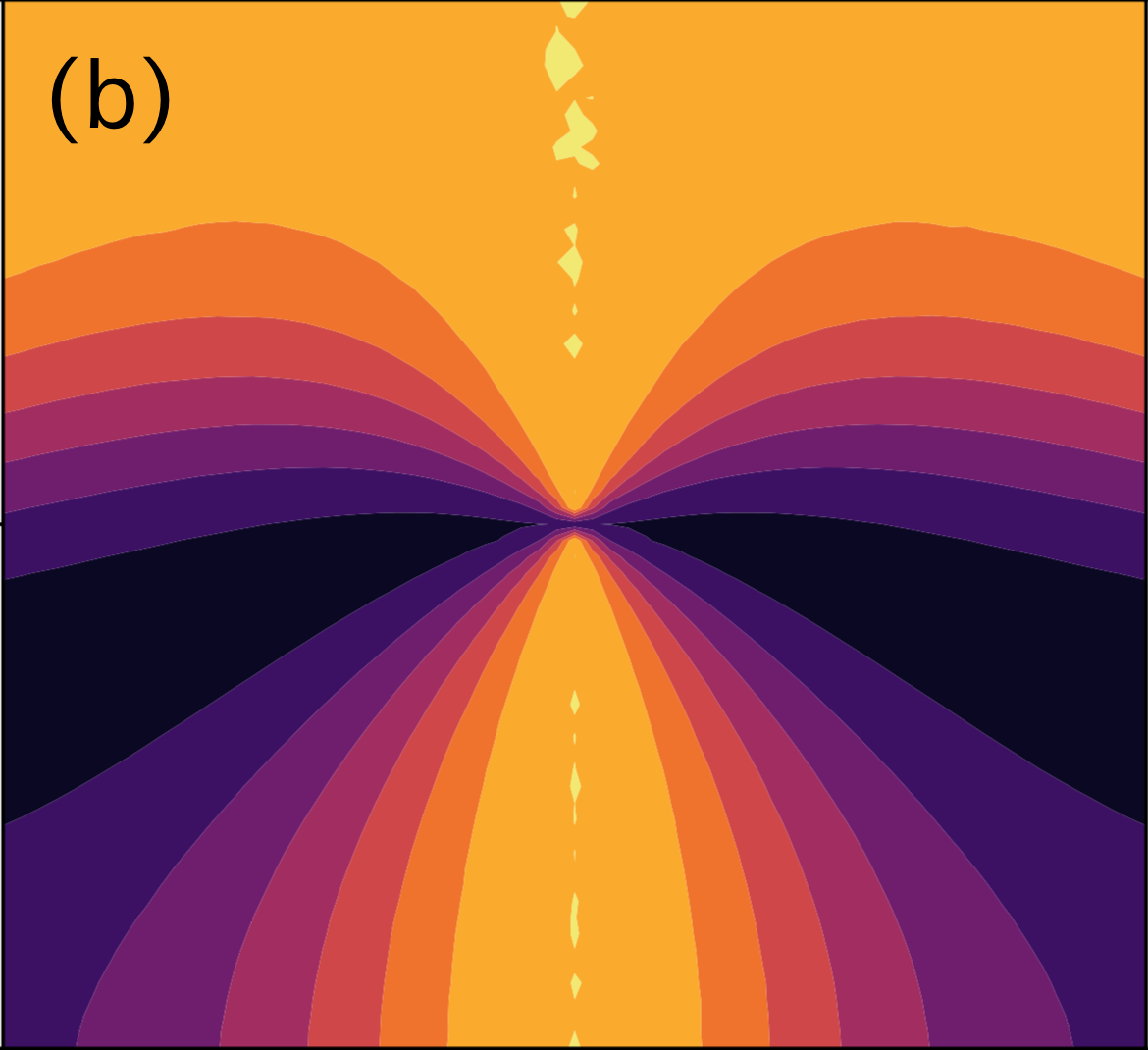}\label{sfig:pp_t14}
  \end{subfigure}
  \begin{subfigure}{0.8\linewidth}
    \includegraphics[width=\textwidth]{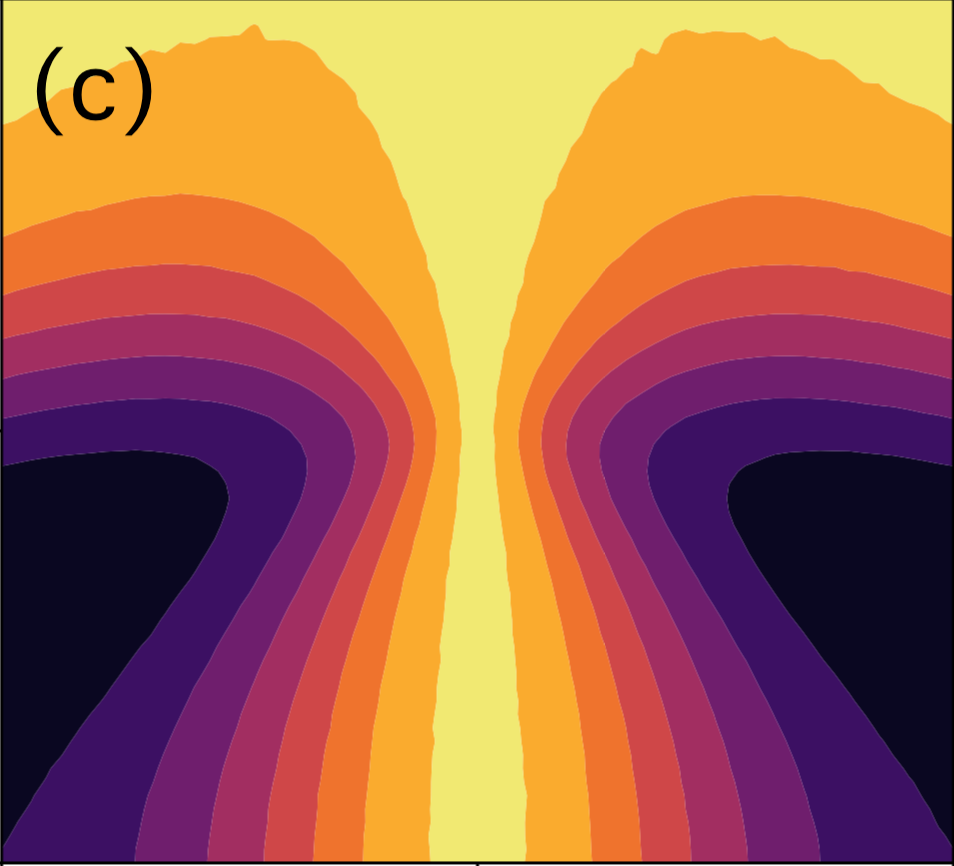}\label{sfig:pp_t18}
  \end{subfigure}
\end{minipage}
\begin{minipage}[h]{.39\linewidth}
  \centering
  \begin{subfigure}{\textwidth}
    \vspace{-2cm}
    \includegraphics[width=\textwidth]{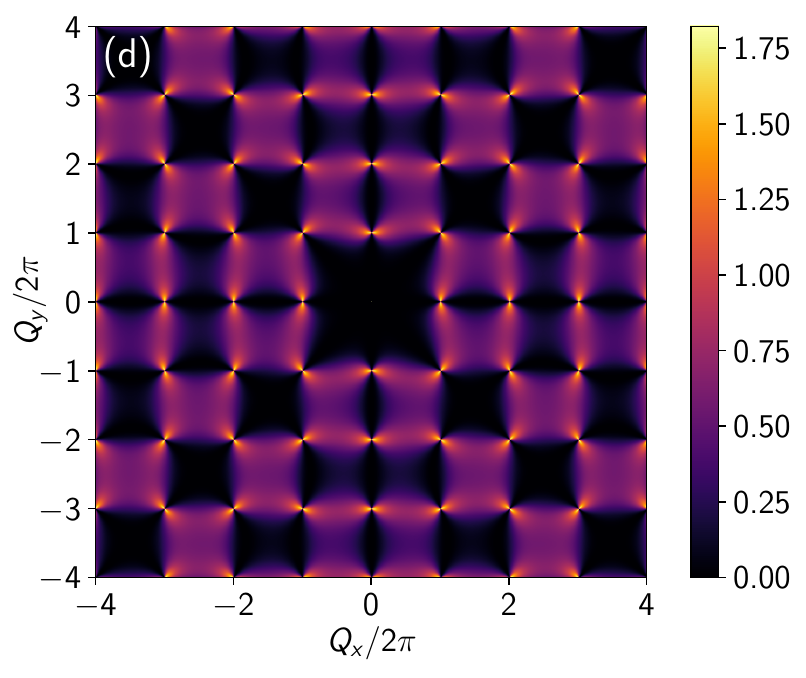}
\end{subfigure}
\end{minipage}
\caption{\raggedright{(a) $S^{\alpha \beta}({\bf Q})$ ($L=128$) projected transverse to ${\bf Q}$ at $T = 0.22$.  (b,c) Pinch points at $ T = 0.22 \approx T_{\rm KT}$ and $ T = 0.29 > T_{\rm KT}$ respectively. (d) Transverse projection of $S^{\rm L}({\bf Q})$ at $T=0.29$ with `longitudinal pinch points'.}}
\end{figure}

We retrieve this phenomenology for the Coulomb fluid on a square lattice in Fig. 6, by projecting our structure factor tensor transverse to the `scattering' vector ${\bf Q}= {\bf G} + {\bf q}$.  Just below $T_{\rm KT}$, (Fig 6a), the fields are predominantly solenoidal and the pattern is almost identical to that observed experimentally in artificial spin ice~\cite{Perrin, Ostman}. The pinch points arise because of the broken rotational invariance in the low temperature regime, where the charge concentration is low and lattice fields are purely solenoidal~\cite{paradox}. Fig. 6b, c shows how, as the system is heated through the BKT transition, the pinch points broaden, as the excitation of deconfined charges restores rotational invariance on length scales longer than the screening length. Such pinch point rounding is observed in spin ice experiments~\cite{paradox}, while they remain sharp in theoretical models, reflecting the induced nature of the monopole charges coming from underlying classical dipoles.  Consequently, the observation of broadened pinch points in very large artificial spin ice arrays would be a signature of magnetic charge that is deconfined and fully screened on all long length scales within the system. However any such effect would go beyond a model of classical dipoles.

Construction of the transverse projection of $S^{\rm L}$ is also of interest, because (Fig. 6d), this contains `anti pinch points' -- `bow-ties' (finite intensity perpendicular to ${\bf Q}$) rather than `hour-glasses'(finite intensity parallel to ${\bf Q}$) -- reminiscent of some antiferromagnets. Such a re-construction from neutron scattering data on three dimensional spin ice materials would result in bow-tie patters strongly resembling those observed in the model pyrochlore antiferromagnets that were highly influential in the early days of frustrated magnetism~\cite{Zinkin}.

It is a pleasure to thank Y. Levin for a useful correspondence, A. Alastuey, B. Canals, T. Dauxois, D. McMorrow and T Roscilde for useful discussions and the following for financial support: EPSRC, UCL, the Leverhulme Trust, ANR grant FISICS, the IUF (Roscilde), the ENS de Lyon and the National Science Foundation under Grant No. NSF PHY-1748958 at KITP.

\pagebreak

\pagebreak

\appendix

\section{Details of the Simulations}

The program is written in Fortran 2008 with OpenMPI used for parallelisation and FFTW for the Fourier transforms of the electric field. In addition to an irrotational and harmonic part, the algorithm~\cite{MR} introduces a freely-fluctuating rotational field, which maintains the thermodynamics of the system because the partition function factorises. 

Three main field updates are used. (1) A field link update which combines charge creation, annihilation and movement. Flux $ E_i \rightarrow E_i \pm Q/\epsilon $ is added to, or subtracted from a randomly chosen field link, which is equivalent to adding or subtracting a unit of charge from one end of the field link and subtracting or adding it at the other end (Fig.~A1). 
\begin{figure}[h]
  \centering
  \includegraphics[width=0.48\textwidth]{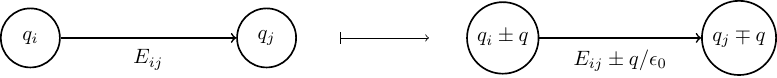}\label{}
  \caption{A field link update.}
\end{figure}

(2) Addition or subtraction of flux $ \Delta $ around a randomly chosen plaquette of field links (Fig.~A2); this allows for relaxation of the total field via sampling of the solenoidal (rotational) degrees of freedom. (3) Addition or subtraction of $ \bar{E}_{\mu} \rightarrow \bar{E}_{\mu} + L \frac{Q}{L^d \epsilon_0} n $ to a given component $ \mu $ of the the harmonic mode of the field is proposed, which corresponds to the change in the harmonic mode arising from a single charge winding around the system once in the $\mu$-direction. This results~\cite{Faulkner,Callum} in a grand canonical energy change of $Q L \left( \frac{q}{2 L \epsilon} \pm \bar{E_{\mu}} \right) $.
\begin{figure}[h]
  \centering
  \includegraphics[width=0.48\textwidth]{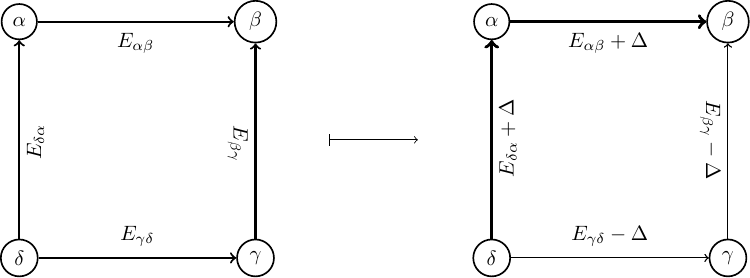}\label{f}
  \caption{A rotational update.}
\end{figure}

All three updates are proposed and accepted or rejected by the Metropolis algorithm. One Monte Carlo sweep consists of $ N = L^2 $ field link updates, $2N$ rotational updates and $N$ harmonic updates.
The grand canonical simulations begin with vacuum; the canonical simulations begin with $n N/2 $ dipole pairs placed randomly throughout the system, with no new charges added or removed as the simulation proceeds.

The simulations were run for 250,000 thermalization sweeps and 500,000 subsequent sweeps for a lattice of linear length $L = 128 \;\; (N = 16384) $, with measurements taken every 20 sweeps.
OpenMPI is used to perform identical simulations with different random seeds, in this case over 32 nodes.
For each measurement various thermodynamic quantities are sampled and the current field configuration is Fourier transformed using the FFTW 2D real-to-complex transform.
After each simulation, the Fourier-transformed correlation tensor \sabq{} was eigendecomposed to extract the longitudinal and transverse eigenvalues, which were then used to construct the longitudinal and transverse field components \slq{} and \stq{}.
The code used can be found at \url{http://github.com/cuamll/mr}.

\section{Two-dimensional electrostatics}

In this appendix we take the electrostatic limit in which the electric field is purely irrotational. In the absence of electrodynamic coupling, an aribitrary solonoidal field can be added without affecting the electrostatic correlations, as discussed in the text.  
The electrostatic energy
\begin{equation}
U=\frac{\epsilon_0}{2}\int |{\bf E}^{\rm L}({\bf r})|^2 d^3 r, 
\end{equation}
includes both the Coulomb energy of interaction and the ``self-energy'' of the particles, $U_{\rm self}=Nu_{\rm self}$ -- the electrostatic energy required to establish the set of independent charges.  
The Coulomb energy is the difference between the total energy and the sum of the self energies and corresponds to the change in the field amplitudes on generating a correlated charge configuration
\begin{equation}
U_{\rm c}=U-U_{\rm self}. \label{Uc}
\end{equation}
As $U$ is quadratic in the field strengths one can define a pairwise Coulomb interaction, $u_c(r)$ as the difference in energy for a system of two charges $q_1$ and $q_2$, separated by distance $r$ and infinity: 
\begin{equation}
u_c(r)=U_2(r)-U_2(L)
\end{equation}
where $L$ is the system size taken to the thermodynamic limit yielding $U_2(L\rightarrow\infty)=2u_{\rm self}$.

The chemical potential
$\mu$ is the energy required to place an isolated charged particle in the system. By convention $\mu$  is defined to be negative if the energy cost for placing a particle is greater than zero.  The total ``Landau'' energy of a charge system is then $U_{\rm c}-\mu N$. 
If the the energy is purely electrostatic, $\mu=-u_{\rm self}$, but one could also have a non-electrostatic contribution, the single particle core energy $\epsilon$ such that 
\begin{equation}
-\mu  =u_{\rm self}+\epsilon.
\end{equation}
It then follows:
\begin{equation}
U_{\rm c}-\mu N= U+\epsilon N.\label{Field-E}
\end{equation}

The Coulomb interaction $u_c(r)$ is found by sovling Poisson's equation. In three dimensions (3D), 
$u_{\rm c}(r)= \frac{q_1q_2}{4\pi \epsilon_0 r}$, which is non-confining, falling to zero at large $r$.  
Two dimensions (2D), on the other hand, is rather special as both the self energy and the Coulomb interaction for charges separated by distance $r$ diverge logarithmically with system size:
\begin{equation}
u_{\rm c}(r)=- \frac{q_1q_2}{2\pi\epsilon_0} \left[\ln(r)-\ln(L)\right]\label{Uc2D}.
\end{equation}
As a consequence, both $U_{\rm c}$ and $\mu$ are singular in the thermodynamic limit. However, these singularities cancel in the Landau energy, $U_{\rm c}-\mu N$ 
allowing for  an extensive energy function for a finite concentration of particles. 
One can circumvent the singularities by considering the reference state as a nearest neighbour neutral charge pair rather than an isolated particle.
For this we define a dipolar self energy
\begin{equation}
u_{\rm self}^{\rm dip}=2u_{\rm self}+u_{\rm c}(a),\label{self-dip}
\end{equation}
and a Coulomb interaction relative to nearest neighbour pair separation:
\begin{equation}
u_{\rm c}^{\rm 2D}=u_{\rm c}(r)-u_c(a)= -\frac{q_1q_2}{2\pi\epsilon_0} \ln\left(\frac{r}{a}\right),
\end{equation}
both of which are well defined.
From from this one can construct a 2D Coulomb energy
\begin{equation}
U_{\rm c}^{\rm 2D}=U_{\rm c}-\frac{N}{2}u(a)=U-\frac{N}{2}u_{\rm self}^{\rm dip}.
\end{equation}

We can now define a 2D chemical potential which relates the energy cost of introducing (half) a nearest neighbour pair of charges, rather than an isolated free charge
\begin{equation}
-\mu^{\rm 2D}=\frac{1}{2}u_{\rm self}^{\rm dip}+\epsilon.
\end{equation}
Finally we can re-write equation (\ref{Field-E}) for the 2D case in terms of well defined quantities 
\begin{equation}
U_{\rm c}^{\rm 2D}-N\mu^{\rm 2D}=U+ N\epsilon. 
\end{equation}
Setting the single particle core energy, $\epsilon=0$ we arrive at the pure electrostatic problem announced in the text. Note that $\mu^{2D}$ is often referred to as the core energy for the nearest neighbour pair (see for example ~\cite{ZH}), which should not be confused with $\epsilon$, the single particle core energy defined in the text (in the field theoretic work of Lee and Teitel, \cite{LeeTeitel} their parameter $u$ corresponds to $u= -\epsilon$ here).

\section{Comparison with the ZH form}

For regimes (ii) and (iii), defined in the main text, ZH~\cite{ZH} derived a thermodynamic limit formula for the correlation function in a low density or fugacity approximation. This translates to 
$
\epsilon_q = 1+ \kappa^2  \left(1 - J_{\rm A}(q a)\right)/(-\Delta_q)
$
with 
\begin{equation}\label{ZH}
J_{\rm A}(q a) = \frac{2 \nu'}{\Gamma[\nu'+1]}\left(\frac{q^2 a^2}{4}\right)^{\nu'/2} K_{\nu'}\left(\sqrt{q^2a^2}\right),
\end{equation}
where $\nu' = \frac{2 \pi \epsilon_0 Q^2}{2kT} - 1$ and $K_{\nu'}$ is a modified Bessel function of the second kind. 
A comparison of this expression with the simulated data is shown in Fig.~B1. The cusp-like ZH form is qualitatively correct at, and below, $T_{\rm KT}$ and can be projected onto the data by a linear transformation $S^{\rm L}(q) \rightarrow m S^{\rm L}(q) + c$  where $m,c$ are fitting parameters.
\begin{figure}[htb!] \centering
\includegraphics[width= 0.45\textwidth]{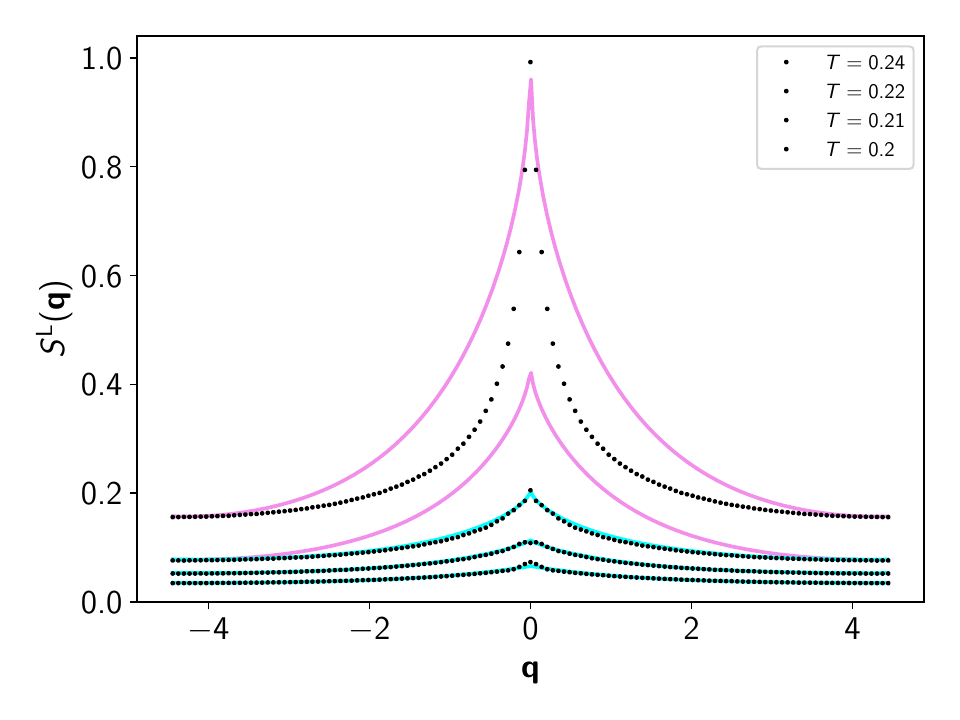}\label{}
\caption{Simulated $ S^{\rm L}(\mathbf{q})$ (black points) compared with ZH theory (magenta line, no fitted parameters) and ZH theory with rescaled peak (cyan line, two fitted parameters; the ZH function is rescaled and then a constant term added). }
\end{figure}

\section{Finite Size Scaling}

Considering  $T_{\rm KT} < T \lesssim T_2$, EK predicted that in the thermodynamic limit, there are two regimes: a small-$q$ regime with `classical' Debye-H\"uckel like correlations and a larger-$q$ power-law regime with  exponent $2/\nu(T)$ (see main text). We analysed $S(q)$ data for system sizes $L = \sqrt{N} = 16, 32, 64, 128, 256$, and confirmed a quantitative agreement with the EK power law prediction for all system sizes, suggesting only very small finite size corrections in this regime. In contrast, the behaviour of $S(q)$ in the classical small-$q$ regime was indicative of very large finite size corrections.  To illustrate this dichotomy, we show here an analysis of the data at two particular $q$ values: $q^{\ast} = 0, 0.55$, chosen to represent the classical and power law regimes respectively. 

\begin{figure}[htb!] \centering
\includegraphics[width= 0.6\textwidth]{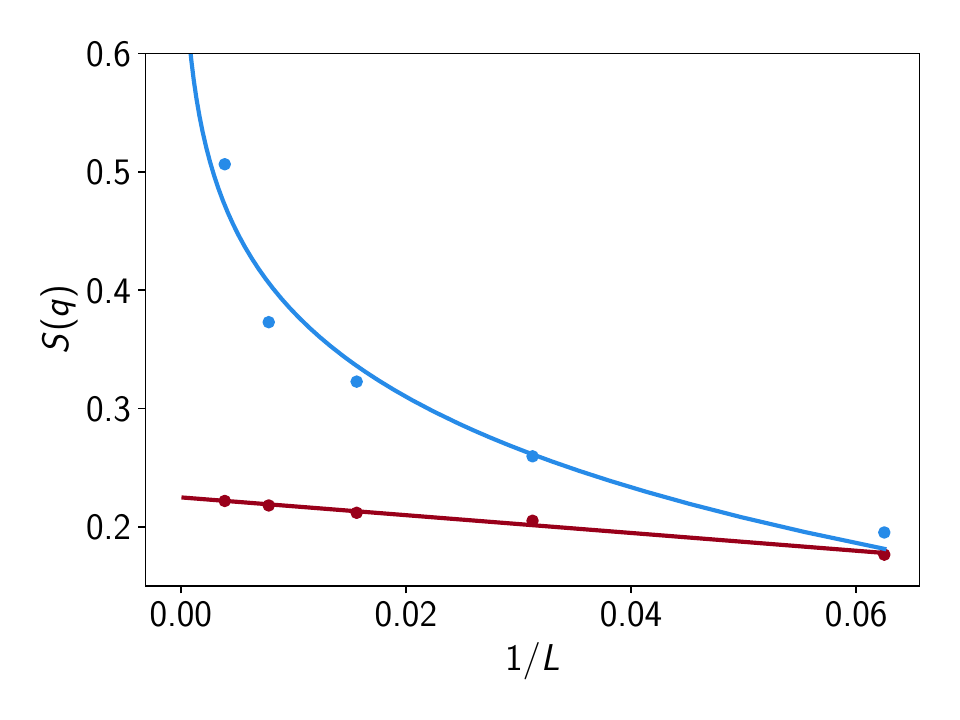}\label{}
\caption{{$S^{\rm L}(q^{\ast})$ at $T = 0.23$ and $q^{\ast} = 0, 0.55$ (blue, red points respectively) as a function of $1/L$. Corresponding lines are fits to Eqs. \ref{eqpower}, \ref{eqline} respectively.} }
\end{figure}

In the power law regime the data $S_{q^{\ast}}(1/L, T)$ was found (Fig. C1)  to fit to the line: 
\begin{equation}\label{eqline}
S_{q^{\ast}}(1/L, T) = m(T)(1/L) + c(T), 
\end{equation}
with the fitted parameter $m(T)$ approaching zero as $T\rightarrow T_{\rm KT}$ and $c(T)$ most likely remaining finite in the same limit: see Fig. C2. Hence the finite size corrections to EK's power law regime are small and consistent with central-limit theorem scaling ($\sim\sqrt(1/N)$). 

\begin{figure}[b!] \centering
\includegraphics[width=0.6\textwidth]{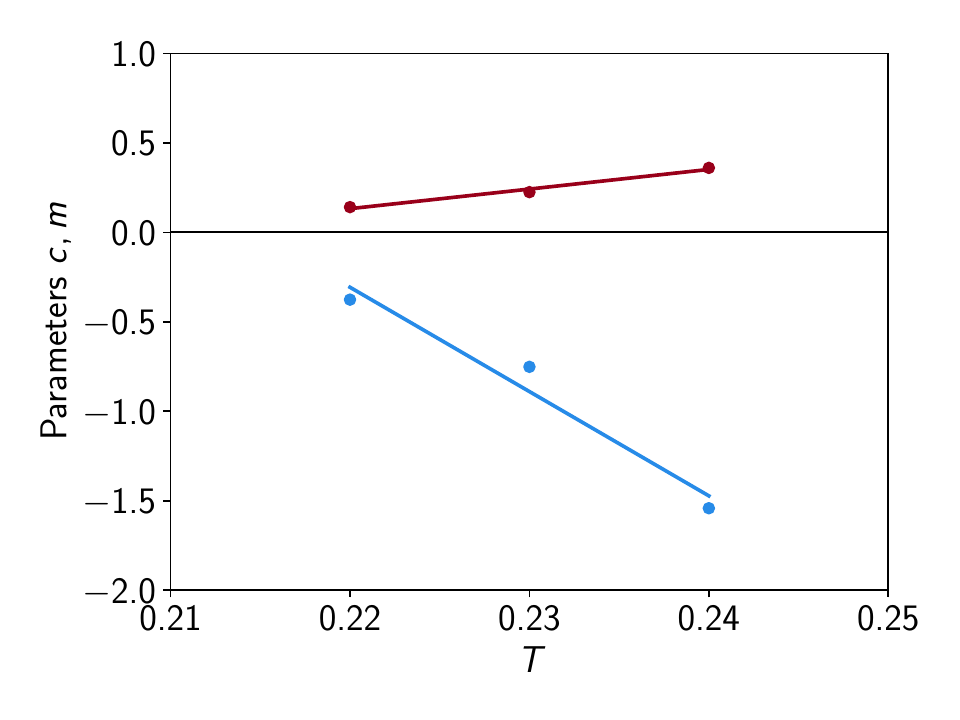}\label{}
\caption{ Temperature dependence of fitted parameters $m$ (blue) and $c$ (red) in Eq. \ref{eqline} fitted to $q^{\ast} = 0.55$ data in the power law regime (lines are guides to the eye).}
\end{figure}

In the classical regime, data at $T  = 0.22,0.23$ for all system sizes and $T = 0.24$ for $L< 256$ could be 
adequately described (Fig. C3) by the formula:
\begin{equation}\label{eqpower}
S_{q^{\ast}}(1/L, T) = 2 \pi T - \alpha(T) (1/L)^{\zeta(T)}, 
\end{equation}
with the fitted amplitude $\alpha(T)$ varying slowly with temperature, and the fitted exponent $\zeta(T)$ linearly approaching zero as $T \rightarrow T_{\rm KT}$, while remaining of order $1/\nu(T)$: see Figs. 6. 

\begin{figure}[ht!] \centering
\includegraphics[width=0.6\textwidth]{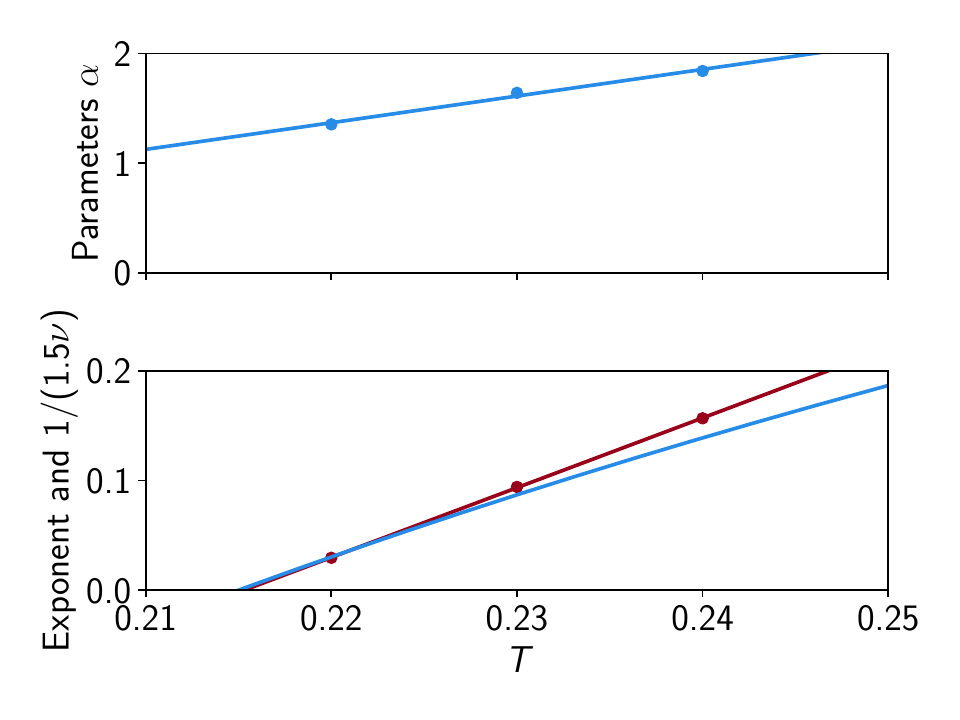}\label{}
\caption{Temperature dependence of fitted parameters $\alpha$ (blue, upper) and $\zeta$ (red, lower) in Eq. \ref{eqpower} fitted to $q^{\ast} = 0$ data in the classical regime (lines are linear fits). The lower blue curve is $2/3\nu(T)$.}
\end{figure}

At small $q$, and at $T = 0.24, 0.25$, the expected rounding and cut off of the  power law of Eqn. \ref{eqpower} at the SL value $S(0) = 2 \pi T$ starts to become visible at small $q$ (see Fig. C4).  Hence, with a power law rounded at small $q$, the $1/L$ dependence of $S(q)$ to a large extent mirrors its $q-$dependence at $1/L = 0$.  However it is a  noteworthy trend that the `anomalous' regime in $q$ has `classical' scaling in $1/L$, while the `classical' regime in $q$ has `anomalous' scaling in $1/L$. 

At temperatures well above $T = 0.25$, our finite simulations are essentially at the thermodynamic limit for all system sizes and temperatures and the SL condition is everywhere obeyed (see e.g. data in Fig. 2a, main text).

\begin{figure}[htb!] \centering
\includegraphics[width= 0.6\textwidth]{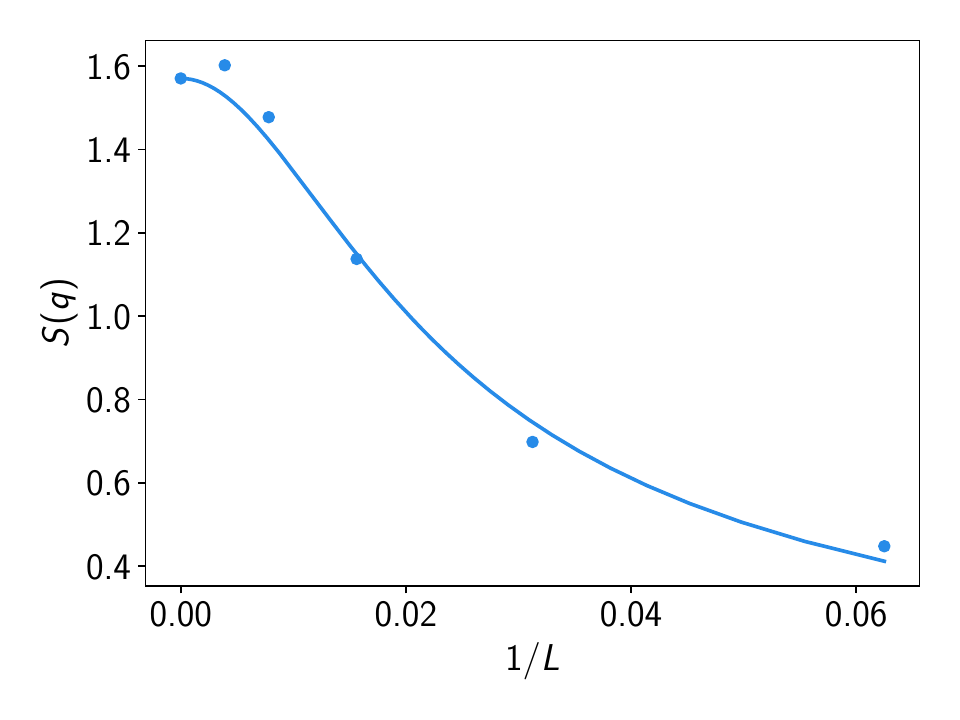}\label{}
\caption{$S^{\rm L}(q^{\ast})$ at $T = 0.25$ and $q^{\ast} = 0$ (points). The blue curve is $1.57/(1 + 3640 L^{-2})^{0.49}$ where the parameters were obtained in a free fit: agreement with the SL result $S(0) = \pi/2$ is confirmed here. }
\end{figure}
 

\begin{thebibliography}{99}
\bibitem{foot} Refs. \cite{Minnhagen, Young} interpreted the poor electrolyte as a gas of coexisting monopoles and dipoles, but the non-analytic properties are more consistent with unbounded dipolar, or multi-scale, pairs.
\bibitem{fh} A harmonic component, confined to $q=0$ for periodic boundaries, is not relevant here.
\end{thebibliography}
\end{document}